\documentclass[twocolumn]{aastex631}

\newcounter{ionstage}
\renewcommand{\ion}[2]{\setcounter{ionstage}{#2}%
  \ensuremath{\mathrm{#1\,\scriptstyle\Roman{ionstage}}}}
 
\newcommand{\hii}{H\thinspace{\sc ii}}

\newcommand\Te{\ensuremath{T_{\mathrm{e}}}}

\received{XXX}
\revised{YYY}
\accepted{ZZZ}

\shorttitle{About Metallicity Variations in the Local Galactic ISM}
\shortauthors{Esteban et al.}
\graphicspath{{./}{figures/}}

\begin{document}

\title{About Metallicity Variations in the Local Galactic Interstellar Medium}

\correspondingauthor{C\'esar Esteban}
\email{cel@iac.es}

\author[0000-0002-5247-5943]{C. Esteban}
\affiliation{Instituto de Astrof\'isica de Canarias (IAC), E-38205 La Laguna, Spain}
\affiliation{Departamento de Astrof\'isica, Universidad de La Laguna, E-38206 La Laguna, Spain}

\author[0000-0002-6972-6411]{J. E. M\'endez-Delgado}
\affiliation{Instituto de Astrof\'isica de Canarias (IAC), E-38205 La Laguna, Spain}
\affiliation{Departamento de Astrof\'isica, Universidad de La Laguna, E-38206 La Laguna, Spain}

\author[0000-0002-6138-1869]{J. Garc\'ia-Rojas}
\affiliation{Instituto de Astrof\'isica de Canarias (IAC), E-38205 La Laguna, Spain}
\affiliation{Departamento de Astrof\'isica, Universidad de La Laguna, E-38206 La Laguna, Spain}

\author[0000-0002-2644-3518]{K. Z. Arellano-C\'ordova}
\affiliation{Department of Astronomy, The University of Texas at Austin, 2515 Speedway, Stop C1400, Austin, TX 78712, USA}



\begin{abstract}

In this paper we discuss and confront recent results on metallicity variations in the local interstellar medium, obtained from observations of {\hii} regions and neutral clouds of the Galactic thin disk, and compare  them with recent high-quality metallicity determinations of other tracers of the chemical composition of the interstellar medium as B-type stars, classical Cepheids and young clusters. We find that the metallicity variations obtained for these last kinds of objects are consistent with each other  and with that obtained for {\hii} regions but significantly smaller than those obtained for neutral clouds. We also discuss the presence of a large population of low-metallicity clouds as the possible origin for large metallicity variations in the local Galactic thin disk. We find that such hypothesis does not seem compatible with: (a) what is predicted by theoretical studies of gas mixing in galactic disks, and (b) the models and observations on the metallicity of high-velocity clouds and its evolution as they mix with the surrounding medium in their fall onto the Galactic plane. We conclude that  that most of the evidence favors that the chemical composition of the interstellar medium in the solar neighborhood is highly homogeneous.

\end{abstract}



\section{Introduction} \label{sec:intro}


The determination of the spatial variations of metallicity in the interstellar medium (ISM) across the Galactic disk is a fundamental constraint for understanding the importance of mixing processes, abundance gradients and the chemical evolution of the Galaxy \citep[i.e.][]{Edmunds75, RoyKunth95}. The presence of significant metallicity variations in the local Galactic ISM has been a matter of debate. Pioneer theoretical works devoted on estimating the time scale and efficiency of the mixing processes of heavy elements ---and O in particular--- in the ISM of galactic disks or in the Milky Way in particular, predicted very small variations \citep[i.e.][]{Edmunds75, RoyKunth95}, several times smaller than those reported in early observational works. For example, studies on the radial gradients of the O abundance in the Galactic disk as those of \citet{shaver83} or \citet{FichSilkey91} were consistent with variations up to a factor two (0.3 dex) at spatial scales of $\sim$1 kpc. Later studies of the Galactic abundance gradient based on radio or far infrared observations found a scatter of the O/H ratio around the gradient fit of about 0.16–0.20 dex or even higher \citep[e.g.][]{Quireza06, Rudolph06}. Large variations were also reported in early works on stellar abundances in B-type stars as that by \citet{Rolleston94}, who found variations of about 0.7 dex in stars in clusters separated by distances of about 1 kpc. However, over time, thanks to the improvement in the depth and quality of observations and spectrum analysis techniques, the dispersion in the distribution of chemical abundances and metallicities has decreased considerably, indicating that much of this scatter ---which was commonly interpreted as intrinsic--- was actually largely due to observational and methodological errors. In the last decade,  there is growing evidence that {\hii} regions \citep[e.g.][]{Esteban18, Arellano-Cordova20, Arellano-Cordova21} and   Population I objects \citep[O and B-type stars, classical Cepheids or young clusters, i.e.][]{Simon-Diaz10, NievaPrzybilla12, Braganca19, Genovali14, Luck18, Donor20}  show fairly smaller variations in their chemical abundance distribution, especially once corrected for the radial abundance gradient.

However, there is no lack of evidence to the contrary. Very recently, \citet{DeCia21} reported the existence of metallicity variations of up to a factor ten in the vicinity of the Sun, based on the study of neutral clouds along the line of sight of 25 O- or B-type stars located in or very close to the Galactic plane and within 3 kpc. They measure an average metallicity of $-$0.26 dex with respect to the solar one (55 per cent solar) finding values ranging between $-$0.76 dex and $+$0.26 dex, where $\sim$ 3/4 of the lines of sight show subsolar metallicities.  \citet{DeCia21} interpret the finding of low-metallicity variations as the product of the falling of high-velocity clouds of pristine gas onto the Galactic disk. If real, the metallicity distribution they observe would imply that the low-metallicity accreting gas does not efficiently mix into the ISM.

In this paper, we present and discuss a compilation of  high-quality metallicity determinations for {\hii} regions and other young population objects to discuss their metallicity distribution and constrain the amplitude of possible spatial variations. 

In Section~\ref{sec:variations} we present results on O abundance variations determined from the analysis of the spectra of Galactic and extragalactic {\hii} regions. In Section~\ref{sec:neutral} we present and discuss recent results on metallicity variations in neutral clouds of the Galactic thin disk and the Magellanic Clouds and compare them with results for {\hii} regions. In Section~\ref{sec:popI} we compare the results for {\hii} regions and neutral clouds with metallicity variations determined from data for different local Galactic Population I objects. In Section~\ref{sec:mixing}, we discuss the likelihood of the existence of a large population of low-metallicity clouds in the local Galactic thin disk. Finally, in Section~\ref{sec:conclusions}, we summarize our main conclusions.

\section{Metallicity variations in galactic disks from {\hii} region observations} \label{sec:variations}

\subsection{Galactic {\hii} Regions}
\label{sec:galhii}

The most recent and accurate determination of the radial abundance gradients of the ionized gas phase of the ISM in the Milky Way is by \citet{Arellano-Cordova20, Arellano-Cordova21}. Due to the characteristics of the sample and the unprecedented quality of their observations and analysis methodology, the results of that study are a very significant improvement over previous studies. \citet{Arellano-Cordova20, Arellano-Cordova21} use spectra of 42 {\hii} regions located at Galactocentric distances ($R_{\text G}$) from 4 to 17 kpc. Among the advantages of that work we can highlight: (a) the use of very deep spectra taken with 8-10m telescopes; (b) the application of the same methodology and updated atomic data for the analysis of a large number of spectra; (c) the direct determination of electron temperature ({\Te}) for all objects; (d) the use of revised distances based on Gaia Data Release 2 (DR2) parallaxes \citep{Gaiadr2, Bailer-Jones18} of the stars associated with the nebulae and (e) a careful selection of the most appropriate ionization correction factor (ICF) scheme for each chemical element, propagating the uncertainty of this magnitude in the determination of total abundances. \citet{Arellano-Cordova20, Arellano-Cordova21} estimate that the mean difference between the individual O abundances and that obtained from the linear fit of the radial O abundance gradient for the corresponding $R_{\text G}$ of each object is 0.07 dex, a value consistent with the typical uncertainties of the  determination of the O/H ratio in individual objects (the median is 0.08 dex). Considering this result, \citet{Arellano-Cordova20, Arellano-Cordova21} conclude that O is fairly well mixed in the Galactic ISM, discarding the presence of significant metallicity variations, at least in the quadrant of the Milky Way covered by their sample of {\hii} regions. 

\begin{figure}[t!]
\epsscale{1.18}
\plotone{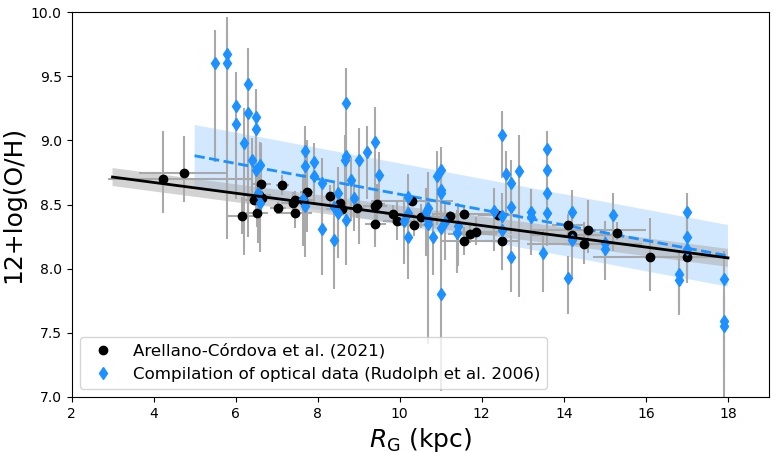}
\caption{Comparison between the O/H ratios and gradient fit obtained by \citet[][black circles and continuous line]{Arellano-Cordova21} and those obtained from the compilation of optical data by \citet[][blue diamonds and dashed line]{Rudolph06}. The coloured bands centred on the gradient lines represent the dispersion of the data points around the fit in each case. 
\label{fig:rudolph}}
\end{figure}

Figure~\ref{fig:rudolph} illustrates the improvement of the quality of the radial oxygen abundance gradient determined by \citet[][black circles and continuous line]{Arellano-Cordova21} with respect to that derived by \citet[][blue diamonds and the blue dashed line]{Rudolph06}, one of the most widely used references of the Galactic O abundance gradient. \cite{Rudolph06} use a compilation of abundances determined from optical spectra where {\Te} is obtained using different methods: the same optical spectrum, from radio observations (not cospatial, usually corresponding to the whole nebulae) or estimated indirectly by other means. The coloured bands centred on the gradient lines represent the dispersion of the individual abundances around the gradient fit in each case; the grey area corresponds to the data by \citet{Arellano-Cordova21} and the blue one to those by \cite{Rudolph06}. The dispersion obtained from the data by \cite{Rudolph06} is about 0.24 dex, four times larger than the dispersion reported by \citet{Arellano-Cordova21}.

\subsection{Extragalactic {\hii} Regions}
\label{sec:extragalhii}

The most recent abundance determinations in {\hii} regions of the Magellanic Clouds by \citet{Dominguez-Guzman22} also indicates their remarkable chemical homogeneity. The work is based on very deep echelle spectroscopy taken with UVES spectrograph at the 8m VLT telescope and uses  a methodology and atomic dataset very similar to those of \citet{Arellano-Cordova20, Arellano-Cordova21} for the Milky Way. All the nebulae have excellent direct determinations of {\Te}. Attending to observational evidence, dwarf irregular galaxies like the Magellanic Clouds are expected to be chemically homogeneous and not to exhibit radial abundance gradient \citep[e.g.][]{kobulnicky97, Croxall09}.
In this sense, \citet{Dominguez-Guzman22} find remarkable similar abundances in all objects inside each galaxy. They obtain a weighted mean of 12 + log$_{10}$(O/H) of 8.01 and 8.37 for SMC and LMC, respectively, with standard deviations of 0.02-0.03 dex, of the order of the median of the individual O abundance error, 0.02 dex. \cite{Dominguez-Guzman22} make the exercise of compiling optical spectroscopical data from previous works of the same {\hii} regions they observed and recalculate their O/H ratios following the same methodology. Thus, they obtain similar mean O/H values but larger standard deviations, of the order of 0.06 dex for both galaxies. This result demonstrates the importance of the quality of the data for diminishing dispersion.

Many works devoted on the determination of metallicities and abundance gradients in external galaxies at different redshifts are based on the application of the so-called strong-line methods to estimate the metallicity of the ionized gas. These methods are used when the temperature-sensitive auroral CELs can not be measured in the spectra of {\hii} regions. i.e. {\Te} can not be derived and the direct method to determine abundances can not be applied. It is well-known that metallicities estimated from strong-line methods are less accurate than those derived from the direct method \citep[e.g.][]{Peimbert17, Perez-Montero17,Arellano-Cordova20b}. This is because they are affected by significant systematic biases depending on the lines used by the different methods and intrinsic dispersion inside a given method \citep[e.g.][]{Lopez-Sanchez12, Maiolino19}. 

\citet{DeCia21}, arguing the existence of large metallicity variations in the Milky Way and other galaxies, cite the results of two works dedicated on estimating the metallicity distribution of a large number of {\hii} regions along the disks of nearby spiral galaxies. Those works analyze optical integral field spectroscopy obtained from the PHANGS-MUSE survey \citep{Kreckel19} and MaNGA and MAD galaxy samples \citep{Wang21}. \citet{DeCia21} remark the finding of variations of metallicity of the order of 0.5 dex in the case of the data of \citet{Wang21} or between 0.2 and 0.3 dex in the case of \citet{Kreckel19}. Both works apply strong-line methods to estimate the O/H ratio. In particular, \citet{Wang21} use the N2S2H$\alpha$ empirical relation by \citet{Dopita16} and the S-calibration estimator by \citet{Pilyugin16}. Although the metallicity values given by both methods are quite similar, the dispersion given by the N2S2H$\alpha$ relation is nearly twice of that obtained using the S-calibration. Despite of this, \citet{DeCia21} only discuss the result obtained from the first method. Moreover, \citet{DeCia21} cite the work by \citet{Kreckel19} just to say that they find metallicity variations of 0.2-0.3 dex from the radial gradients; however, \citet{Kreckel19} indicate explicitly that the finding of a very low scatter of the metallicity (0.03--0.05 dex) at any given radius is the main conclusion of their paper and an indication of the high degree of mixing in all the galactic disks they studied.  

\citet{Bresolin11} demonstrated that the use of high-quality, direct determinations of O abundance in {\hii} regions decreased appreciably the scatter in the O abundance along the disk of M33. This author obtained a scatter of 0.06 dex for that galaxy, much lower than the 0.21 dex obtained by \citet{Rosolowsky08} using the $R_{\rm 23}$ strong-line method and optical spectra of lower signal-to-noise ratio. Other extensive works devoted to derive direct determinations of the O/H from data of many {\hii} regions in nearby spiral galaxies, like those of the CHAOS (CHemical Abundance Of Spirals) project, find values of the scatter with respect to the gradient fit between 0.04 and 0.10 dex for M101, NGC~628, NGC~2403, NGC~3184, and NGC~5194 \citep{Croxall16,Berg20,Rogers21}. Also using direct abundance determination, \citet{Esteban20} obtain scatter values of 0.10 and 0.04 dex, for M31 and M101, respectively. In all these cases, the values of the scatter are of the order or only slightly larger than the typical observational uncertainty of the individual O/H determinations \citep[see discussion in][]{Esteban20}. It is important to note that the aforementioned values of the scatter are fairly consistent with the amplitude of the abundance variations found in Galactic {\hii} regions as well as the level of inhomogeneity of the ISM predicted by hierarchical models of star formation, which \citet{Elmegreen98} estimates to be of the order of 0.05 dex

\begin{figure}[t!]
\epsscale{1.4}
\plotone{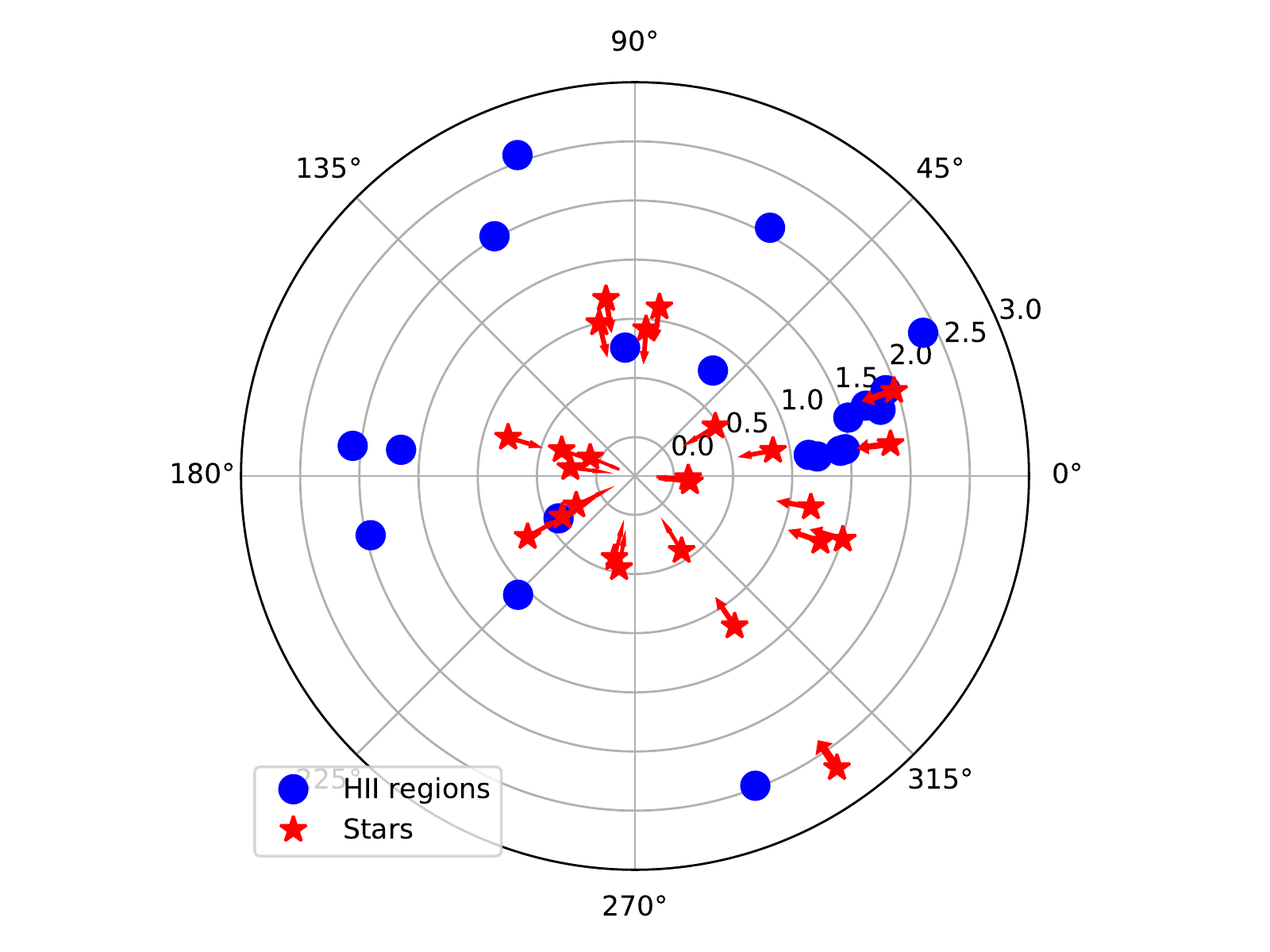}
\caption{Spatial distribution around the Sun and onto the Galactic plane of the O- and B-type stars (red stars) used as lines of sight of absorption systems by \citet{DeCia21}. The red stars have arrows because their position correspond to upper limits of the true distance of the neutral clouds. The Sun is at the center and each concentric circle represents a heliocentric distance of 0.5 kpc. The blue dots represent the position of the {\hii} regions of \citet{Arellano-Cordova20, Arellano-Cordova21} located at heliocentric distances less than 3 kpc. \label{fig:distribution}}
\end{figure}

\section{Metallicity variations in the local Galactic disk from neutral clouds observations} \label{sec:neutral}

As it is said in Section~\ref{sec:intro}, \citet{DeCia21} reported the existence of large metallicity variations in the solar vicinity. Those authors estimate the metallicity of neutral clouds located along the line of sight of several O- or B-type stars located in the Galactic plane within 3 kpc of the Sun. In Figure~\ref{fig:distribution}, we show the spatial distribution around the Sun and onto the Galactic plane of the lines of sight used by \citet{DeCia21} and the {\hii} regions observed by \citet{Arellano-Cordova20, Arellano-Cordova21} located at heliocentric distances $<$ 3 kpc. The figure includes 25 stars and 20 {\hii} regions. The heliocentric distances of the stars used by \citet{DeCia21} have been taken from the Gaia DR2. It must be taken into account that the positions of the stars correspond to upper limits to the true distance to the neutral clouds whose metallicity \citet{DeCia21} estimate.  The heliocentric distances to the {\hii} regions have been taken from \citet[][]{Mendez-Delgado22} and are derived from the Gaia Early Data Release 3 \citep[hereinafter EDR3,][]{Gaiadr3} parallaxes applying the probabilistic approach of \citet{Bailer-Jones21}. 

\begin{figure}[ht!]
\epsscale{1.15}
\plotone{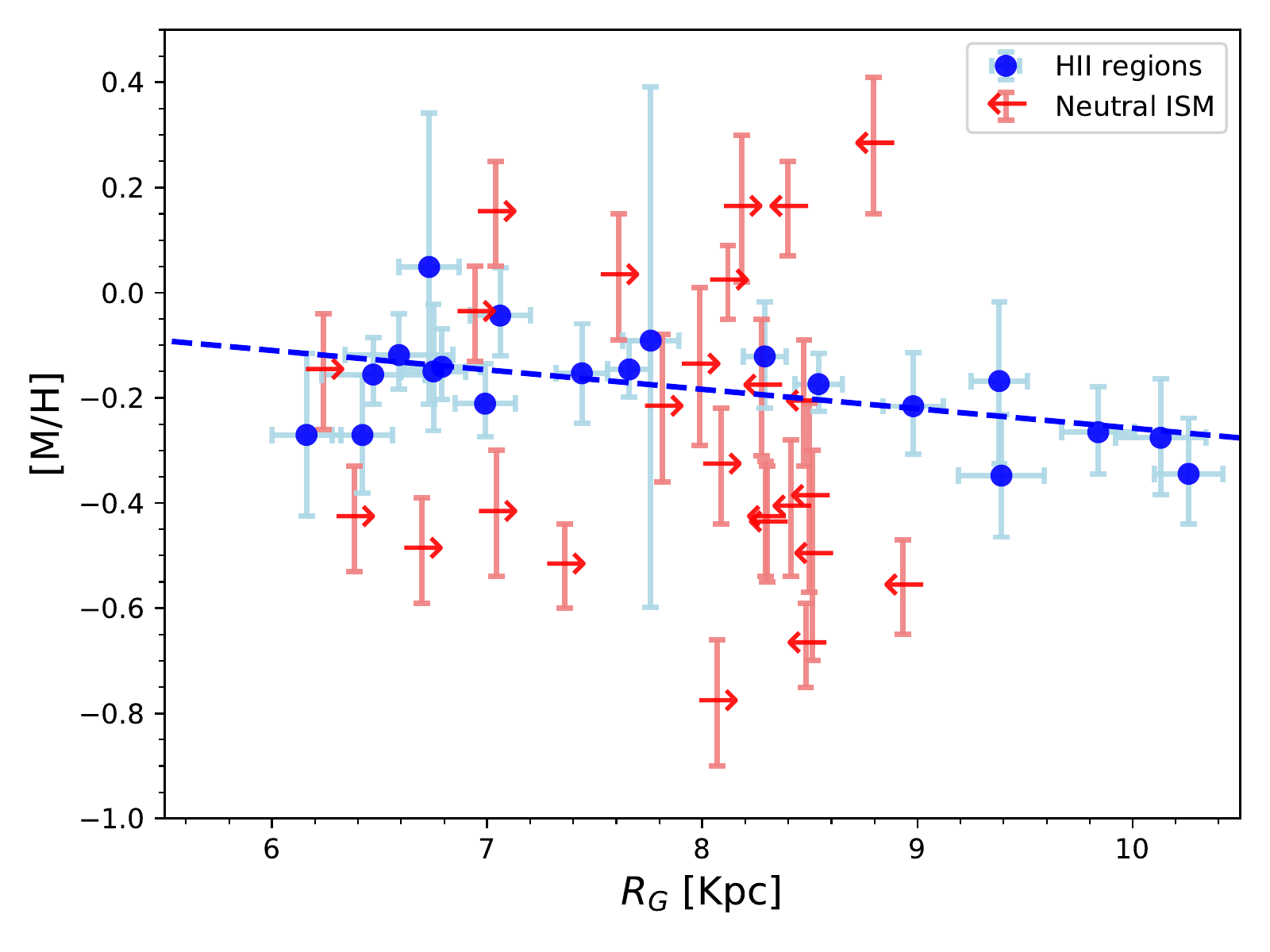}
\caption{Metallicity with respect to the solar value as a function of $R_{\text G}$ for neutral clouds in the line of sights observed by \citet{DeCia21} (red arrows) and {\hii} regions of \citet{Arellano-Cordova20, Arellano-Cordova21} located at heliocentric distances less than 3 kpc (blue dots). The determination of [M/H] in the case of the {\hii} regions is described in the text. The red arrows indicate the upper or lower limits of the $R_{\text G}$ of the neutral clouds. The dashed blue line represents the lineal fit to the {\hii} region abundances represented in the figure. \label{fig:gradient}}
\end{figure}

In Figure~\ref{fig:gradient}, we compare the metallicity of the objects represented in Figure~\ref{fig:distribution} as a function of their corresponding $R_{\text G}$. We represent the relative metallicity of the objects with respect to the solar one, in the form: [M/H] = $\log_{10}$(M/H) $-$ $\log_{10}$(M/H)$_\odot$. In Figure~\ref{fig:gradient}, the lines of sight of \citet{DeCia21} are represented as red arrows, whose sense indicates if their given $R_{\text G}$ correspond to a lower or upper limit. The metallicities of the {\hii} regions are obtained from the O/H ratios given by \citet{Arellano-Cordova21} subtracting 12$+\log_{10}$(O/H)$_\odot$ = 8.69, the value of the solar O abundance used as reference by \citet{DeCia16, DeCia21}, that corresponds to the photospheric O/H ratio recommended by \citet{Asplund09}. As we can see, the metallicity distributions of the two kinds of objects represented in Figure~\ref{fig:gradient} are completely different. It is remarkable the extremely large spread of [M/H] values found by \citet{DeCia21} between 8 and 9 kpc, a range of $R_{\text G}$ that includes the maximum and minimum metallicities they report. \citet{DeCia21} explicitly say that their observational data do not show evidence of a radial metallicity gradient, arguing that the narrow range of $R_{\text G}$ covered by their line of sights may be the reason for that. However, the {\hii} regions represented in Figure~\ref{fig:gradient} do show a gradient slope of $-$0.037$\pm$0.019 dex kpc$^{-1}$ (represented by a dashed blue line in Figure~\ref{fig:gradient}), consistent with the one obtained by \citet{Arellano-Cordova21} of $-$0.042$\pm$0.009 dex kpc$^{-1}$ for their whole sample that covers $R_{\text G}$ values between 4 and 17 kpc. 

\begin{deluxetable*}{ccccccccc}
\tablenum{1}
\tablecaption{Mean metallicities and variations in different young population objects\label{tab:variations}}
\tablewidth{0pt}
\tablehead{
 & \multicolumn2c{Neutral Clouds} & & \multicolumn3c{} & Classical & Young \\
 & \multicolumn2c{DC21} & {\hii} Regions & \multicolumn3c{B Stars} & 
 Cepheids & Clusters \\
& Relat. Met. & F$^*$ Met. & AC21 & \multicolumn2c{NP12} & B19 & L18 & D20 
}
\startdata
Distance range & \multicolumn2c{$d_{\text{Hel}}$$<$3 kpc} & 
$d_{\text{Hel}}$$<$3 kpc & \multicolumn2c{$d_{\text{Hel}}$$<$0.5 kpc} & 
 $d_{\text{Hel}}$$<$3 kpc & $d_{\text{Hel}}$$<$3 kpc & $R_{\text{G}}$=5--11 kpc \\ 
Elements & \multicolumn2c{Several} & O & O &  & O & Fe & Fe \\
No. Objects & \multicolumn2c{25} & 20 & \multicolumn2c{20} & 9 & 45 & 10 \\
Mean [M/H] & $-$0.17 & $-$0.24 & $-$0.18 & 0.07& 0.05 & 0.01 & 0.04 & $-$0.04 \\
$\sigma$([M/H]) & 0.29 & 0.22 & 0.09 & 0.05 & 0.02 & 0.06 & 0.09 & 0.10 \\
Weighted Mean [M/H] & $-$0.49 & $-$0.48 & $-$0.18 & 0.06 & 0.06 & $-$0.05 & 0.02 & $-$0.14 \\
Weighted $\sigma$([M/H]) & 0.27 & 0.23 & 0.06 & 0.05 & 0.02 & 0.07 & 0.07 & 0.11 \\
Median [M/H] error & 0.11 & 0.12 & 0.08 & 0.09 & 0.10 & 0.07 & 0.09 & 0.02 \\
$\left[\text{M/H}\right]_{\text{min}}$ & $-$0.78 & $-$0.69 & $-$0.35 & $-$0.05 & 0.01 & $-$0.13 & $-$0.10 & $-$0.28 \\ 
$\left[\text{M/H}\right]_{\text{max}}$ & 0.28 & 0.11 & $-$0.04 & 0.15 & 0.08 & 0.08 & 0.28 & 0.12  \\
$\left[\text{M/H}\right]_{\text{max}}$$-$$\left[\text{M/H}\right]_{\text{min}}$ & 1.06 & 0.80 & 0.31 & 0.20 & 0.07 & 0.21 & 0.38 & 0.40 \\
\multicolumn9c{Quantities defined with respect to the gradient fit determined in each reference} \\
$\left[\text{M/H}\right]$ scatter & -- & -- & 0.06 & -- & -- & 0.03 & 0.06 & 0.06 \\
$\Delta\left[\text{M/H}\right]$ & -- & -- & 0.27 & -- & -- & 0.16 & 0.43 & 0.22 \\
\enddata
\tablecomments{DC21: \citet{DeCia21}; AC21: \citet{Arellano-Cordova21}; NP12: \citet{NievaPrzybilla12}; B19: \citet{Braganca19}; L18: \citet{Luck18}; D20: \citet{Donor20}. }
\end{deluxetable*}

In Table~\ref{tab:variations}, we compare the metallicity variations found by \citet{DeCia21} with those obtained by \citet{Arellano-Cordova21} for {\hii} regions as well as other Population I objects. We give two different set of values for \citet{DeCia21} because they use two approaches for estimating neutral cloud metallicities, the `relative method' that provides dust-corrected abundances for individual metals, introduced by \citet{DeCia16}, and the `F*' method that gives the total metallicity, developed by \citet{Jenkins09}. In the table, we give the mean and standard deviation of [M/H] ($\sigma$([M/H])) as well as these same magnitudes weighted by the inverse of the square of the uncertainty of each individual metallicity determination. In addition, we include in Table~\ref{tab:variations} the median of the uncertainty of [M/H] as well as the minimum and maximum [M/H] values of each data set and the difference of both quantities ($\text{[M/H]}_{\text{max}}-\text{[M/H]}_{\text{min}}$). In the last two lines of the table, we include the mean difference between the individual [M/H] values with respect to the gradient fit --what we call [M/H] scatter--, and the maximum amplitude of the differences of [M/H] measured  above and below the gradient fit ($\Delta$[M/H]) but only for those data sets where a gradient has been calculated. In each column of Table~\ref{tab:variations} we also indicate the element (O or Fe) that has been used as proxy of metallicity in each reference. The corresponding [M/H] for each object has been determined subtracting the solar O/H or Fe/H ratios adopted by  \citet{DeCia16, DeCia21}. In the case of Fe, those authors use 12$+\log_{10}$(Fe/H)$_\odot$ = 7.475, the mean of the photospheric and meteoritic solar abundances recommended by \citet{Asplund09}. 

\begin{figure}[ht!]
\epsscale{1.15}
\plotone{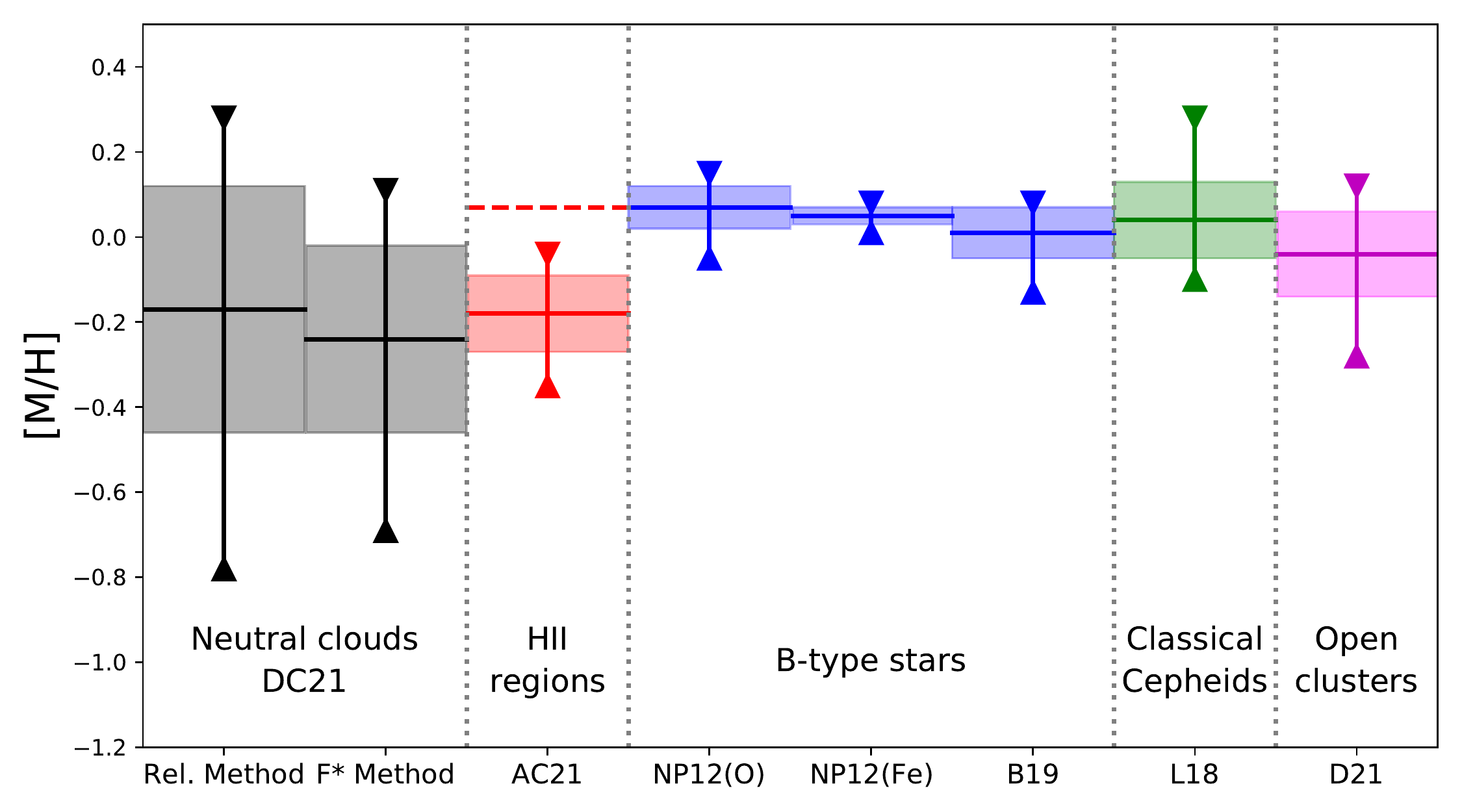}
\caption{Metallicity variations in neutral clouds, {\hii} regions and other Population I objects in the local Galactic disk. Horizontal solid lines represent mean metallicities, the height of the colored rectangles represent  $\sigma$([M/H]), and the down- and up-pointing triangles represent the maximum and  minimum values of [M/H], respectively, for each kind of object. Black and grey symbols represent data for neutral clouds by \citet[][DC21]{DeCia21}, which include calculations using the relative and F* methods. Red symbols correspond to data for {\hii} regions obtained by \citet[][AC21]{Arellano-Cordova21}. Blue symbols represent metallicities for B-type stars derived from O and Fe abundances by \citet[][NP12]{NievaPrzybilla12} and by \citet[][B19]{Braganca19} from O abundances. Green symbols correspond to data for classical Cepheids \citep[][L18]{Luck18}. Magenta symbols represent metallicities for a sample of open clusters \citep[][D21]{Donor20}. The horizontal dashed red line represents the average [M/H] value obtained for the {\hii} regions studied by AC21 and considering $t^2 > 0$ \citep{Mendez-Delgado22}.\label{fig:comparison}}
\end{figure}

In Figure~\ref{fig:comparison}, we show some quantities included in Table~\ref{tab:variations} for a more graphic comparison. We represent the mean of the [M/H] values --not the weighted mean-- (horizontal lines), as well as their $\sigma$([M/H]) (colored rectangles) and the maximum and  minimum values of [M/H] of the different data sets. The length of the solid vertical lines delimited by triangles correspond to the [M/H]$_{\text{max}}-$[M/H]$_{\text{min}}$ quantity given in Table~\ref{tab:variations}. 

In Table~\ref{tab:variations} and Figure~\ref{fig:comparison}, we can note that the mean metallicity of neutral clouds and {\hii} regions is fairly similar. However, it seems remarkable that since the mean and weighted mean of [M/H] are almost the same for {\hii} regions, they differ by a factor of 0.24 or 0.32 dex (depending on the method for correcting for dust depletion used) in the case of neutral clouds. This indicates that the locus of neutral clouds with the best metalliticy determinations are displaced to low [M/H] values. From Table~\ref{tab:variations} and  Figures~\ref{fig:gradient} and \ref{fig:comparison} it is evident that $\sigma$([M/H]) and [M/H]$_{\text{max}}-$[M/H]$_{\text{min}}$ values are dramatically different for neutral clouds and {\hii} regions. It is also remarkable that while $\sigma$([M/H]) is of the order of the median uncertainty of the individual determinations for {\hii} regions, both quantities differ in the case of neutral clouds. This indicates that the metallicity distribution of neutral clouds is further from a normal distribution than that of {\hii} regions. As expected, for {\hii} regions, the scatter of [M/H] with respect to the linear fit to the gradient is smaller than $\sigma$([M/H]).


\subsection{Dust and O abundance determination in {\hii} regions} 
\label{sec:dust}

In ionized nebulae analysis, the O/H ratio is the proxy for the metallicity, as Fe/H is for stellar abundances. Much of our understanding of the chemical composition and evolution of galaxies comes from the analysis of the spectra of {\hii} regions, and one can easily check that dust depletion has not been considered a big concern when determining metallicities from the O abundance in these objects \citep[e.g.][]{Peimbert17, Perez-Montero17, Maiolino19}. This contrast with the statement of \citet{DeCia21} saying that ``the effects of dust depletion have prevented a deeper investigation of the ISM metallicity". In our opinion, this comment underestimates the huge amount of work done on determining abundances of the ionized-gas phase of the ISM from the analysis of {\hii} region optical spectra in both, the Galactic and extragalactic domains.  Furthermore, \citet{DeCia21} say that the determination of the metallicty of the Orion Nebula ---the reference object for the gas-phase metallicity of the solar neighbourhood--- is debated ``ranging from 1/10 solar metallicity for collisionally excited lines, and in general for refractory elements such as Fe, Mg and Si to slightly supersolar". This statement is not correct. Firstly,  It is a matter of fact that all specialists working on ionized nebulae know that refractory elements like Mg, Si or Fe are heavily depleted onto dust grains under the prevailing conditions of {\hii} regions and, therefore, they are never used for estimating metallicities in these objects. Secondly, the O/H abundance ---and therefore the metallicity--- of the ionized gas of the Orion Nebula is very well constrained. Numerous deep and detailed spectroscopical studies in different parts of the Orion Nebula give completely consistent results \citep[i.e.][]{Osterbrock92, Esteban04, Mendez-Delgado21} with differences of the order of 0.03 dex. The dust depletion factor of O is considered small in {\hii} regions. In fact, \citet{mesadelgado09} estimated that the fraction of O embedded in dust grains in the Orion Nebula should be about 0.12 dex at most. In a more general study of Galactic and extragalactic {\hii} regions, \citet{Peimbert10} estimate that the depletion fraction of O abundance increases with increasing O/H, from about 0.08 dex in the most metal poor {\hii} regions to about 0.12 dex in metal-rich ones. \citet{Peimbert10} propose that 0.10 dex is a good representative value for objects of solar metallicity; however, that value of the possible value of O depletion is of the order of the typical uncertainties of the O/H ratio in most {\hii} regions, therefore, difficult to be noted.

\section{Metallicity variations from local Population I objects} 
\label{sec:popI}

In addition to neutral clouds and {\hii} regions, Table~\ref{tab:variations} and Figure~\ref{fig:comparison} also include metallicity variations derived for different kinds of Population I objects belonging to the local Galactic thin disk: early B-type stars, classical Cepheids and young open clusters. Due to their  short lifetimes, all these stellar objects are still located not far from their birth place and are unaffected by enrichment processes due to stellar nucleosynthesis or depletion onto dust grains. Therefore, their metallicity distribution should be representative of the present-day one of the local ISM. The metallicity values of the different objects included in Table~\ref{tab:variations} or Figure~\ref{fig:comparison} are calculated with respect to the solar O or Fe abundances adopted by \citet{DeCia16, DeCia21}.

The metallicities of \citet{NievaPrzybilla12} included in Table~\ref{tab:variations} and Figure~\ref{fig:comparison} are based on photospheric abundances of O and Fe of 20 bright, single, sharped lined and chemically inconspicuous early B-type stars located in the solar neighbourhood, at heliocentric distances $<$500 pc. Their observational data consist on high-resolution and high-S/N spectra taken with different spectrographs that are analysed applying non-LTE line-formation computations in a self-consistent way. In Table~\ref{tab:variations} and Figure~\ref{fig:comparison}, we also present data obtained by \citet{Braganca19}, the most recent study of Galactic radial abundance gradients using young main-sequence OB stars. They use a sample of 31 stars located in the outer Galactic disk and perform a non-NLTE analysis of high-resolution spectra obtained with the MIKE spectrograph on the Magellan Clay 6.5-m telescope at Las Campanas Observatory. \citet{Braganca19} find a linear metallicity gradient with a somewhat stepper slope than that obtained by \citet{Arellano-Cordova20, Arellano-Cordova21} for {\hii} regions but consistent with the recalculation made by \citet{Mendez-Delgado22} assuming $t^2>0$ (see Section~\ref{sec:adft2}). Since the sample of \citet{Braganca19} covers stars located at $R_{\text G}$ between 8.4 and 15.6 kpc, we only consider those overlapping the area covered by the observations of \citet{DeCia21}, corresponding to stars with heliocentric distances $<$ 3 kpc. As we can see in Table~\ref{tab:variations} and Figure~\ref{fig:comparison}, the mean [M/H] values obtained from the data used for B-type stars, are clearly larger than those obtained for neutral clouds or {\hii} regions, but their $\sigma$([M/H]) and [M/H]$_{\text{max}}-$[M/H]$_{\text{min}}$ values, are far more consistent with those obtained from {\hii} regions than from neutral clouds.
There are other metallicity determinations in early type stars available in the literature. From determinations of the O/H ratio for a sample of 13 B-type stars of the Ori OB1 association, \citet{Simon-Diaz10} finds a $\sigma$([M/H]) and [M/H]$_{\text{max}}-$[M/H]$_{\text{min}}$ of 0.04 dex and 0.10 dex, respectively. In fact, this author highlights the high degree of chemical homogeneity of the Ori OB1 stars. Finally, \citet{Lyubimkov15} derive stellar abundances of 22 B-type stars with heliocentric distances $\leq$ 600 pc. Those authors find metallicity variations somewhat larger than the other cited references: $\sigma$([M/H]) = 0.13 dex and [M/H]$_{\text{max}}-$[M/H]$_{\text{min}}$ = 0.39 dex, but still far from the larger variations shown by neutral clouds. 

Classical or Population I Cepheids are variable stars with masses between 4 and 20 M$_{\odot}$ \citep{Turner96}, which correspond to ages $\lesssim$300 Myr \citep{Bono05}. From a LTE analysis of high-resolution spectra, \citet{Luck18} determine Fe abundances for several hundreds of Galactic classical Cepheids located at $R_{\text G}$ values from 3 to 18 kpc, the largest compilation of abundances of classical Cepheids to date. \citet{Luck18}  finds a linear gradient fit with a slope very similar to that obtained for {\hii} regions. Of the sample studied by that author, we only consider the results for a subsample of 52 Cepheids with the highest quality abundance determinations based on observations covering five or more phases of the variability period, of those we select the 45 objects located in the Galactic area covered by the observations of \citet{DeCia21}, having heliocentric distances $<$ 3 kpc. Table~\ref{tab:variations} and Figure~\ref{fig:comparison} show that the metallicity variations  --parameterized with $\sigma$([M/H]) and [M/H]$_{\text{max}}-$[M/H]$_{\text{min}}$-- found for local classical Cepheids are similar to those obtained for {\hii} regions. \citet{Korotin14} determined the radial O abundance gradient from data of several hundreds of Cepheids, most of them Type II Cepheids, old low-mass Population II giants, with ages of about several Gyr. Those authors find $\sigma$([M/H]) and [M/H]$_{\text{max}}-$[M/H]$_{\text{min}}$ values of 0.15 and 0.42 dex, respectively, somewhat larger than  the variations obtained for classical Cepheids and {\hii} regions --as expected because their much larger ages-- but still clearly smaller than the numbers obtained by \citet{DeCia21} for neutral clouds.  

In the last column of Table~\ref{tab:variations} and in Figure~\ref{fig:comparison}, we include the metallicity parameters for  open clusters of the OCCAM (Open Cluster Chemical Abundances and Mapping) survey published by \citet{Donor20}. This work makes use of infrared-based spectroscopic data from the APOGEE-2/SDSS-IV DR16 \citep{apogee2_DR16} of more than a hundred of open clusters. In our case we have selected a subsample containing the 10 youngest clusters --with ages $<$400 Myr-- flagged as ``high quality" in the database and located in the interval of $R_{\text G}$ between 5 and 11 kpc, consistent with the area covered by the observations by \citet{DeCia21}. The values of $\sigma$([M/H]) and [M/H]$_{\text{max}}-$[M/H]$_{\text{min}}$ for these clusters, 0.10 and 0.40 dex, respectively, are consistent with those of the other Population I objects and {\hii} regions included in Table~\ref{tab:variations}.

A final aspect we want to highlight from the data gathered in Table~\ref{tab:variations} is the remarkable consistency of the values of the scatter of [M/H] around the gradient fit and $\Delta$[M/H]$_{\text{max}}$ obtained for the different kinds of Population I objects. This strongly suggests that the true degree of homogeneity of the local ISM should be not far from those numbers. 

In Section~\ref{sec:extragalhii}, we commented the results by \cite{Dominguez-Guzman22}, showing the remarkable homogeneity of the metallicity in the Magellanic Clouds. The almost constant O abundance they find in the LMC contrast with the results obtained for neutral clouds by \citet{Roman-Duval21}. These last authors determine gas-phase metallicities from Hubble Space Telescope UV spectra in the line of sight of 32 massive stars of the LMC, finding that the metallicity increases 0.8 dex from the west to the east side of the galaxy. Considering that the {\hii} regions observed by \cite{Dominguez-Guzman22} cover also both sides of the LMC, the apparent metallicity distribution found by \citet{Roman-Duval21} should be related not to true gas-phase variations but to differences in the fraction of dust depletion along the LMC disk. In fact, this last possibility is also considered by \citet{Roman-Duval21}, who interpret the variation of dust depletion as a likely product of the increased feedback from star formation and the turbulence induced by the collision with the Milky Way. 

\subsection{The abundance discrepancy and metallicity in {\hii} regions} 
\label{sec:adft2}

In Table~\ref{tab:variations} and Figure~\ref{fig:comparison} we can see that the mean metallicity of {\hii} regions is about 0.20 dex lower than that of the rest of  Population I objects. This is related to the fact that the O/H ratio of the Orion Nebula and other {\hii} regions of the solar neighbourhood is lower than solar. The radial gradient of O/H obtained by \citet{Arellano-Cordova21} gives a value of 12 + log$_{10}$(O/H) = 8.50 at the $R_{\text G}$ of the Sun (8.2 kpc), which is 0.19 dex lower than the solar O abundance adopted by \citet{DeCia21}. This offset is also observed in all the other chemical elements whose abundances are derived from CELs in {\hii} regions \citep[e.g.][]{Arellano-Cordova20}, indicating that this difference can not be due to dust depletion ---since it is observed even with a similar amplitude in noble gases such as Ar or Ne, whose depletion into dust grains is highly unlikely---. However, such difference disappears when using recombination lines (RLs) instead of CELs for deriving the O abundance, this is the so-called abundance discrepancy problem. Taking into account the results obtained in different works \citep[see compilation given by][]{Arellano-Cordova20}, the mean abundance discrepancy for O/H in Galactic {\hii} regions is about 0.20 dex. The origin of the abundance discrepancy problem in ionized nebulae is still unsolved, but it may be related to the presence of fluctuations in the spatial distribution of electron temperature \citep[the so-called temperature fluctuations, $t^2$, introduced by][]{Peimbert67} inside {\hii} regions \citep[see][]{Garcia-Rojas07, Esteban18b}.  In a very recent paper, \citet{Mendez-Delgado22} discuss the effect of $t^2$ on the Galactic radial abundance gradients derived from the spectra of the sample of {\hii} regions of \citet{Arellano-Cordova20, Arellano-Cordova21}. Those  authors find that total abundances determined from CELs for all elements at a given $R_{\text G}$ increase 0.2--0.3 dex when $t^2 > 0$ is considered, becoming more consistent with solar values, except in the case of N. The position of the horizontal dashed red line in Figure~\ref{fig:comparison} illustrates the effect of considering an average value of $t^2 > 0$ \citep{Mendez-Delgado22} in the O abundances of the {\hii} regions. In this case, the mean metallicity of {\hii} regions becomes completely consistent with those of the rest of Population I objects.

Finally, it is important to have in mind that, although the O/H ratio derived from {\hii} regions using intensity ratios of CELs may be somewhat lower than those derived from RLs, the net effect does not increase the dispersion in its Galactic distribution, which follows a very tight radial gradient. It rather represents a simple offset in the line fit  \citep[][]{Garcia-Rojas07,Esteban18b,Mendez-Delgado22}. Therefore, the abundance discrepancy problem is not expected to introduce additional scatter in the metallicity derived from the spectra of Galactic {\hii} regions. 


\section{Metallicity variations in the local Galactic disk and gas mixing} \label{sec:mixing}

The comparison of the data compiled in Table~\ref{tab:variations} and  Figure~\ref{fig:comparison} provides a fairly clear picture: the amplitude of  metallicity variations found in {\hii} regions, B-type stars, classical Cepheids and young clusters are similar and relatively small, implying that the ISM is basically chemically homogeneous in the azimuthal direction of the Galactic disk around the Solar neighborhood, consistently with an efficient gas mixing, which is also the prediction by theoretical works \citep[i.e.][]{Edmunds75, RoyKunth95, deAvillez02, Yang12}, where metallicity inhomogeneities are wiped out in the azimuthal direction in less than a Galactic orbital period \citep[i.e.][]{Petit15}. As \citet{RoyKunth95} discussed, there are several mixing mechanisms in the ISM showing different timescales that increase as the the spatial scale increases. At large spatial scales, from 1 -- 10 kpc, cloud-cloud collisions in a shear flow due to differential rotation produce effective mixing in the azimuthal direction at timescales $\leq$10$^9$ yr. For intermediate spatial scales, 0.1 -- 1 kpc, those cloud-cloud collisions in addition to expanding interstellar bubbles --powered by supernova explosions and stellar winds-- are combined with the shear flow to produce mixing with shorter timescales, around 10$^8$ yr, of the order of the half-revolution time of the spiral pattern at the Sun. At smaller spatial scales, 1 -- 1000 pc, turbulent diffusion in ionized, neutral and molecular clouds have different timescales from 10$^7$ to 10$^9$ yr, but fluid instabilities in {\hii} regions may can be as short as 10$^6$ yr. 

Considering that disk rotation is the main agent for mixing in spiral galaxies and the effect of selective loss of metals due to galactic winds, \citet{RoyKunth95} suggested that dwarf galaxies --with weaker rotation fields and lower masses-- should show larger metallicity variations. However, as we discuss in sections~\ref{sec:extragalhii} and \ref{sec:neutral}, this is not confirmed by the observations in dwarf irregular galaxies and in the most recent abundance determinations of {\hii} regions in the Magellanic Clouds by \cite{Dominguez-Guzman22}, who find a highly homogeneous metallicity distribution along their disks, indicating the apparent large efficiency of mixing processes even in dwarf irregular galaxies.

Comparing results for neutral clouds inside the Orion and the Ophiucus areas, \citet{DeCia21} make estimates of the minimum physical scale of the metallicity variations they observe, finding that they are of the order of tens of parsecs or even down to a few parsecs. In Figure~\ref{fig:distribution} we can see a group of eight {\hii} regions rather close to each other located at Galactic longitudes between 6$^\circ$ and 19$^\circ$ and covering heliocentric distances between 1.15 and 1.91 kpc. The mean distance between the {\hii} regions of this group is $\sim$150$\pm$60 pc and their $\sigma$([M/H]) and [M/H]$_{\text{max}}-$[M/H]$_{\text{min}}$ values determined from the data obtained by \citet{Arellano-Cordova20, Arellano-Cordova21} are 0.07 and 0.22 dex. These numbers are very similar to those obtained for the whole subsample of {\hii} regions of \citet{Arellano-Cordova20, Arellano-Cordova21} included in Table~\ref{tab:variations}. This result is consistent with the interpretation that much of the metallicity variations that we find in {\hii} regions may be related to the quality of the observations and uncertainties in abundance determinations.

It is well known that current chemical evolution models explain the presence of radial abundance gradients in spiral galaxies in the context of the inside-out formation paradigm. The galactic disks grow by means of radially dependent gas infall from the halo that drives star formation \citep[i.e.][]{Tosi88, Matteucci89, Boissier99, Chiappini01}. 
This is a strong argument that the Milky Way has accreted --and is still accreting-- low-metallicity gas to maintain its current star-formation activity. \citet{DeCia21} interpret the large metallicity variations they found in the neutral clouds of the thin disk as the product of the falling of high-velocity clouds\footnote{Clouds showing $|v_{\text LSR}| > 100$ km s$^{-1}$ and located in the Galactic halo, further away than 3 kpc of the Milky Way disk \citep{Richter17}.} (HVC) of pristine gas with zero depletion onto the Milky Way. Necessarily, in this scenario, those clouds should not mix efficiently with the ISM until they reach the thin disk in order to maintain their low metallicity. However, the theoretical and observational evidence available does not seem to support that scenario. 

Three-dimensional hydrodynamical simulations of cloud–interstellar medium interactions, predict that mixing between the cloud and its surrounding gas is very efficient, raising the metallicity of the HVC as it crosses the Galactic halo \citep{Gritton14, Henley17, Heitsch22}. Presumably, since the interaction begins long before the clouds slow to the velocity of the Galactic ISM and dissipate, this increase in metallicity should be very important when the HVC reaches the thin disk. In fact, as the most recent calculations made by \citet{Heitsch22} predict, contamination of a HVC traveling through the halo may be so large that the final hydrogen mass may contain a fraction less than 10 per cent of the original cloud material. As \citet{Heitsch22} remark, {\it in all likelihood, metallicity measurements in HVCs will not represent original cloud properties}, and surely they will be much higher when those HVC reach the thin disk. Observed [M/H] values in HVCs, as the Magellanic Stream or Complex C, located at distances of $\sim$50 and 10 kpc, respectively, range between $-$1.0 and $-$0.3 dex \citep{Wakker99, Fox10, Shull11, Richter13, Richter17}. We can see that those metallicities are similar to the ones reported by \citet{DeCia21}, but the distances to the Galactic disk of HVCs are far more larger than the distances to the local neutral clouds observed by \citet{DeCia21}, which are located at heights $\lesssim$ 250 pc above the Galactic plane, $\sim$2/3 of them lying within 100 pc. Moreover, considering that $\sim$90 percent of the accretion rate of the Milky Way has its origin in the Magellanic System \citep{Richter17}, one would expect that the population of neutral clouds coming from such accretion at the  Galactic plane should show metallicities larger than those of the Magellanic Clouds ---[M/H] = $-$0.32 and $-$0.68 for LMC and SMC, respectively \citep{Dominguez-Guzman22}--- due to mixing as they fall and decelerate. 

Intermediate-velocity clouds (IVC) have $|v_{\text LSR}| \approx 30-100$ km s$^{-1}$ and tend to be at heights  between 0.5 and 1 kpc above the Galactic plane \citep{Wakker01}. They are believed to originate in the Galactic Fountain or from HVCs coming from the intergalactic medium slowed down as they fall towards the Galactic plane. IVCs show metallicities between $-$0.3 and 0.0 dex \citep[i.e.][]{Richter01a, Richter01b, Wakker01, Richter17}, higher than those of HVCs. Cloud IV21 --which lies 300 pc above the disk-- is the IVC with the lowest known metallicity. \citet{Hernandez13} measure [M/H] = $-$0.43 $\pm$ 0.12 dex for this object, proposing that IV21 is a decelerated infalling low-metallicity HVC that is mixing with disk gas in the lower Galactic halo. About 1/3 of the sightlines observed by \citet{DeCia21} show metallicities lower than that value despite their lower heights with respect to the Galactic plane. 

Considering all the aforementioned theoretical and observational considerations about the origin, distance distribution, initial metallicities and mixing processes of infalling HVCs and IVCs, we consider unlikely that they may be origin of a significant population of low metallicity neutral clouds in the local Galactic thin disk. 


\section{Conclusions} \label{sec:conclusions}

In this paper we discuss and confront recent results about metallicity variations in the local Galactic ISM obtained from {\hii} regions and neutral cloud observations. While the data by \citet{Arellano-Cordova20, Arellano-Cordova21} indicate that the gas in {\hii} regions is fairly well mixed at a given Galactocentric distance, \citet{DeCia21} report the finding of gas-phase metallicity variations of up to a factor of ten in neutral clouds along the line of sight of massive stars located within 3 kpc around the Sun. We compare these discrepant results with high-quality recent metallicity determinations of other objects that are considered tracers of the chemical composition of the present day ISM: B-type stars, classical Cepheids and young clusters. We find that the metallicity variations obtained for these last kinds of objects are consistent with each other, similar to that obtained for {\hii} regions, but significantly smaller than those obtained for neutral clouds. Moreover, recent direct abundance determinations in {\hii} regions of the Magellanic Clouds and other nearby spiral galaxies obtain metallicity variations of the order of the errors of individual abundance determinations, in agreement to the Milky Way results obtained by \citet{Arellano-Cordova20, Arellano-Cordova21}.

The hypothesis of a large population of low-metallicity clouds in the Galactic thin disk around the Sun proposed by \citet{DeCia21} is also in contradiction with the prediction of theoretical studies of gas mixing. These studies agree on that metallicity inhomogeneities are wiped out in the azimuthal direction of the disk in less than a Galactic orbital period. On the other hand, that scenario does not fit the picture provided by observational results about  metallicty distribution in HVCs falling onto the Galactic plane, as well as models about the mixing of HVCs with the surrounding ISM. We expect that the final products of decelerated HVCs when reaching the Galactic plane should produce a population of clouds with larger metallicities than those found by \citet{DeCia21}. 

The disagreement between the amplitude of the metallicity variations found in neutral clouds and the rest of the objects have two possible explanations: (a) one of the two results are incorrect or (b) if both are correct, the gas mixing in the thin disk is mainly produced just at the onset of star formation, in the transition phase between neutral cloud to {\hii} region. If we consider the first explanation, it is very difficult to explain why four kinds of objects: {\hii} regions, B-type stars, classical Cepheids and open cluster, give similar results ---even similar gradients--- considering the completely different techniques used to derive metallicities in them. Only the results on neutral clouds by \citet{DeCia21} break an otherwise coherent picture, so it is reasonable to think that the problem lies in this latter data set, perhaps related to an incorrect correction for dust depletion. The second possibility, that the gas mixing is produced at the onset of star formation seems rather unlikely. This hypothetical process would require a quite short timescale that will imply a short spatial scale for mixing. Moreover, a process with such limitations would much probably be unable to produce the small metallicity variations we see in all the objects ---apart from neutral clouds--- at scales of several kpc. In any case, if the two results are correct, neutral clouds have very large metallicity variations and {\hii} regions and young stars do not, this would imply that the presence of a significant population of low-metallicity clouds definitely does not have any practical influence on the subsequent composition of the stellar population and, therefore, on the chemical evolution of the Galactic disk.

\begin{acknowledgments}
We acknowledge support from the Agencia Estatal de Investigaci\'on del Ministerio de Ciencia e Innovaci\'on (AEI-MCINN) under grant {\it Espectroscop\'ia de campo integral de regiones \ion{H}{2} locales. Modelos para el estudio de regiones \ion{H}{2} extragal\'acticas} with reference 10.13039/501100011033. JEM-D thanks the support of the Instituto de Astrof\'isica de Canarias under the Astrophysicist Resident Program and acknowledges support from the Mexican CONACyT (grant CVU 602402). JG-R acknowledges support from an Advanced Fellowship under the Severo Ochoa excellence program CEX2019-000920-S. The authors acknowledge support under grant P/308614 financed by funds transferred from the Spanish Ministry of Science, Innovation and Universities, charged to the General State Budgets and with funds transferred from the General Budgets of the Autonomous Community of the Canary Islands by the MCIU.  

\end{acknowledgments}

\bibliography{Esteban_etal_2022}{}
\bibliographystyle{aasjournal}



\end{document}